\UseRawInputEncoding
\documentclass[onecolumn]{revtex4-2}

\usepackage{amsmath,amssymb}
\usepackage[colorlinks,
   linkcolor=blue,
  anchorcolor=blue,
  citecolor=blue]{hyperref}

\usepackage{subfigure}
\usepackage{graphicx}
\usepackage{dcolumn}
\usepackage{bm}
\usepackage[section]{placeins}
\usepackage{xcolor,float}
\usepackage{epstopdf}
\usepackage{booktabs}
\usepackage{threeparttable}
 \usepackage{multirow}
 \usepackage{booktabs}

\begin{document}

\title{Validity of black hole complementarity in the context of generalized uncertainty principle}
\author{ Shurui Wu \footnote {gs.srwu20@gzu.edu.cn}$^{1}$}
\author{ Bing-Qian Wang \footnote {wangbingqian13@yeah.net}$^{2}$}
\author{ Z. W. Long  \footnote {Corresponding author:zwlong@gzu.edu.cn}$^{1}$}
\author {Hao Chen  \footnote {haochen1249@yeah.net}$^{3}$}

\affiliation{$^1$ College of Physics, Guizhou University, Guiyang, 550025, China.\\ $^{2}$ College Pharmacy, Guizhou University of Traditional Chinese Medicine, Guiyang 550025, China. \\$^{3}$ School of Physics and Electronic Science, Zunyi Normal University, Zunyi 563006, China.}

\date{\today}
 \begin{abstract}
Recently, Elias C. Vagenas et al and Yongwan Gim et al studied  the validity of the no-cloning theorem in the context of generalized uncertainty principle (GUP), but they came to conflicting conclusions. With this in mind, we investigate the corrections to the temperature for Schwarzschild black hole in the context of different forms of GUP, and obtain the required energy to duplicate information for the Schwarzschild black hole, it shows that the required energy is greater than the mass of black hole, i.e. the no-cloning theorem in the present of GUP is safe.
\end{abstract}

 \maketitle
{\centering\section { Introduction}}

The nature of Hawking radiation \cite{ret SW} can be comprehended as a semi-classical evaporation of black hole: the mass will decrease with the ongoing thermal emission, then the black hole gets smaller, hotter and radiates faster, finally the black hole breaks down to its eventual disappearance, this process would lead to the information loss paradox \cite{ret SWH}. One of the resolutions for this problem is based on the Hawking radiation carrying the black hole information. In this line of thought,  outside the horizon, the local observer (Bob) could gather the information of the infalling matter state through the Hawking radiation after the Page time \cite{ret DN}, then he decides to dive in the black hole. If the infalling observer (Alice) is crossing the event horizon of the black hole carrying all information of the infalling matter state, and she sends the message to Bob, thus Bob may have the duplicated information which will be a violation of the no-cloning theorem. Regarding this new problem, black hole complementarity has been proposed to reconcile general relativity and quantum mechanics \cite{ret LS}, which means that such a paradox never occurs since the observer inside the horizon is not in the causal past of any observer who measures the information through the Hawking radiation outside the horizon \cite{ret LSU}. Interestingly, a gedanken experiment in the Schwarzschild black hole spacetime shows that the required energy to correlate the observations of both sides of the horizon exceeds the mass of the black hole \cite{ret LS}, which implies that the information has to be encoded into the message with super-Planckian frequency to duplicate it, that is to say, the no-cloning theorem for black hole complementarity is safe for the Schwarzschild black hole.

On the other hand,  it has been claimed that the minimal length defined by Generalized Uncertainty Principle (GUP) \cite{ret AF, ret SD, ret RJ} can play a crucial role in the solution of the Information Loss Paradox \cite{ret NI} and in particular, GUP can prevent the violation of black hole complementarity assuming that the GUP parameter is proportional to the number of fields \cite{ret PC}. Based on the relationship between the required energies and the mass of the black hole, the required energy for Alice to send the message to Bob who jumped into the black hole at the Page time was calculated in the presence of GUP, and compared it to the mass of the black hole, it shows that the GUP improved the black hole complementarity \cite{ret YG}, this result is really amazing. Motivated by this work, Elias C. Vagenas et al investigated the validity of the no-cloning theorem in the presence of linear and quadratic terms of momentum in GUP, they proposed that Bob can simultaneously hold in his hands the information from Alice inside the black hole and the same information which was collected by him while standing outside the stretched horizon, it shows that the no-cloning theorem is safe even in the presence of GUP is inaccurate\cite{ret EC}. Obviously, a discussion of the same type of physics problem leads to two distinct-different conclusions, which is disturbing. Importantly, it seems that the main reason is that they use different forms of GUP, in particular, the latter has one more item (linear term of momentum) than the former, so we have a question, whether is the interaction of the momentum linear term and the momentum quadratic term that invalidates the black hole complementarity? in other words, we intend to identify the reason: whether the more the GUP form contains the different power function of uncertain momentum, the less protective the black hole complementarity owns.

Since the close relationship between the required energy for Alice to send the message and the modified temperature of the Schwarzschild black hole in the context GUP affects the judgment of the conclusion, it implies that it is important to obtain an accurate GUP-corrected temperature of the Schwarzschild black hole. According to the latest literature\cite{ret DU}, regarding to the GUP-corrected thermodynamics of Schwarzschild black hole, Xin-Dong Du et al proposed an improved method and proved that whether the GUP parameter is positive or negative, the improved method can explain the evaporation of black hole more reasonably than the original method. With this in mind, in order to verify our doubts, we intend to obtain an accurate GUP-corrected temperature of the Schwarzschild black hole by using the improved method in the context of the new higher order GUP. Furthermore, regarding the two different results in Ref. \cite{ret YG} and Ref. \cite{ret EC}, by using the improved method, we recalculate the GUP-corrected Hawking temperature and obtain the required energy for Alice to send information.

The remainder of this paper is organized as follows. In Sec.\ref{secII}, we give a brief review of the improved method and derive the expressions for the Page time and the required energy. In Sec.\ref{secIII}, by using the improved method, we calculate the Page time and the required energy in the context of the new higher order GUP. In Sec.\ref{secIV}, we recalculate the corresponding GUP-corrected Hawking temperature and report the new results. Finally, the work is summarized in Sec.\ref{secV}. In this work, we use the natural units $\hbar=c=1$.

{\centering  \section{A brief review of the improved method and Black hole complementarity} \label{secII} }

{\centering  \subsection{ The improved method}}

The well known uncertainty relation defined by position and momentum reads
\begin{equation}
\Delta x \Delta p \geq \frac{1}{2},    \label{eq1} \end{equation}
Based on that, in the presence of strong gravity, one intends to modify this uncertainty principle, and such modification also has support from string theory \cite{ret DA}, thus the result leads to GUP as
\begin{equation}
\Delta x \Delta p \geq \frac{1}{2}\left(1+\alpha l_p^2 \Delta p^2\right),   \label{eq2}     \end{equation}
where $\alpha$ and $l_{\mathrm{P}}=\sqrt{G}$ represent the dimensionless parameter of order unity and the Planck length respectively, obviously, this formalism recovers the Heisenberg uncertainty principle if $\alpha=0$.
The uncertainty range of momentum can be obtained as
\begin{equation}
\frac{\Delta x }{\alpha l_p^2}\left[1+\sqrt{1-\frac{\alpha l_p^2}{(\Delta x)^2}}\right] \geq \Delta p \geq \frac{\Delta x}{\alpha l_p{ }^2}\left[1-\sqrt{1-\frac{\alpha l_p{ }^2}{(\Delta x)^2}}\right],       \label{eq3}   \end{equation}
since the right-hand part of Eq.\eqref{eq3} can reduce to the Heisenberg uncertainty principle if $\alpha=0$, while the left-hand part can not do that, thus Eq.\eqref{eq3} should be rewritten as
\begin{equation}
\Delta p \geq \frac{\Delta x }{\alpha l_p{ }^2}\left[1-\sqrt{1-\frac{\alpha l_p{ }^2}{(\Delta x)^2}}\right],   \label{eq4}    \end{equation}
it should be noted that we assume $\alpha>0$ in this work. Since in string theoretical derivations of GUP, $\alpha$ is essentially the Regge slope parameter $\alpha^{\prime}>0$, related to the string length $\lambda_s$ by $\lambda^{2}_s=\hbar\alpha^{\prime}$.

According to the Ref.\cite{ret DU}, there are two approximation methods to express the uncertainty range of momentum, one is the original method
\begin{equation}
\frac{\Delta x}{\alpha l_p^2}\left(1-\sqrt{1-\frac{\alpha l_p^2}{\Delta x^2}}\right) \approx \Delta p_A, \label{eq5}  \end{equation}
another is the improved method
\begin{equation}
\frac{\Delta x}{\alpha l_p^2}\left(1-\sqrt{1-\frac{\alpha l_p^2}{\Delta x^2}}\right) =  {(\Delta p_G)_{min}}, \label{eq6}  \end{equation}
where $\Delta p_A$ and ${(\Delta p_G)_{min}}$ are corrected uncertainty of the momentum and the corrected minimum uncertainty of the momentum respectively. The only difference between the two methods is whether a minimum limit for the corrected uncertainty of the momentum is contained. Specifically speaking, the original method shows the lower bound of the momentum correction uncertainty is a constant independent of $\Delta x$, i.e. $\alpha>0,\left(\Delta p_A\right)_{\min } \approx \frac{1}{l_p \sqrt{\alpha}},(\Delta x)_{\max } \approx l_p \sqrt{\alpha}$, while the improved method shows the lower bound of the momentum correction uncertainty is a function of $\Delta x$. which implies that the lower bound of the momentum correction uncertainty obtained by the improved method successfully inherits the characteristics of the generalized uncertainty principle while the original method loses most features. Thus, through the corresponding derivation\cite{ret DU}, an approximate expression for the revised differential of the area $d A_G $ is given by
\begin{equation}
d A_G \approx \frac{\left(\Delta A_G\right)_{\min }}{(\Delta A)_{\min }} d A \approx \frac{\left(\Delta p_G\right)_{\min }}{(\Delta p)_{\min }} dA, \label{eq7}  \end{equation}
where ${(\Delta p)_{\min }}$ is the uncorrected minimum uncertainty of the momentum and $A$ represents the area of the black hole horizon. Obviously, the core of the improved method Eq.\eqref{eq7} can be applied to calculate the following various thermodynamic quantities.

{\centering  \subsection{ Black hole complementarity}}

The Schwarzschild black hole metric can be expressed as
\begin{equation} d s^2=-\left(1-\frac{2 G M}{r}\right) d t^2+\left(1-\frac{2 G M}{r}\right)^{-1} d r^2+r^2\left(d \theta^2+\sin ^2 \theta d \phi^2\right)
, \label{eq8}  \end{equation}
where $M$ is the black hole mass and $G$ is gravitational constant. For being convenient to study, we rewrite Eq.\eqref{eq8} in Kruskal-Szekeres coordinates as

\begin{equation} d s^2=-\frac{32 G^3 M^3}{r} e^{-\frac{r}{2 G M}}dUdV, \label{eq9}  \end{equation}
with $U=\pm e^{-\frac{\left(t-r^*\right)}{4 G M}}, V=e^{\frac{\left(t+r^*\right)}{4 G M}}, r^*=r+2 G M \ln \left(\frac{|r-2 G M|}{2 G M}\right)$.

Heisenberg uncertainty principle is given by
\begin{equation}
\Delta X \Delta P \geq 1,    \label{eq10} \end{equation}
$A$ and $T$ are the area and temperature of the black hole identified with $A=4 \pi r_h^2=16 \pi G^2 M^2$  and $T=\frac{f^{\prime}\left(r_h\right)}{4 \pi}=\frac{1}{8 \pi G M}$ for the Schwarzschild black hole, respectively. In order to obtain the Page time for the old black hole \cite{ret DN}, we consider the Stefan-Boltzmann law
\begin{equation}
\frac{d M}{d t}=-A \sigma T^4,   \label{eq13} \end{equation}
where $\sigma$ denotes the Stefan-Boltzmann constant, then the Page time can be obtained from the Stefan-Boltzmann law as\cite{ret YG}
\begin{equation} t_p \sim G^2 M^3,   \label{eq14} \end{equation}
here we suppose that Alice first jumps into the horizon at $V_A$, and then Bob dives the horizon with a record of his measurements of information from the Hawking radiation after the Page time $t_p$ at $V_B$. In order to receive the message from Alice before Bob hits the curvature singularity at $UV=1$, Alice should send the message to Bob before she reaches $U_A=U_B=V_B^{-1}=e^{-\frac{t_p}{4 G M}}$. Therefore, the proper time $ \Delta \tau$ for Alice to send the message to Bob at least at $U_A$, where $\Delta V_A$ is a nonvanishing finite value near $V_A$ for the free-fall \cite{ret LS}. Then, according to Eq.\eqref{eq10}, the required energy $\Delta E$ should demand
\begin{equation} \Delta E \geq \frac{1}{\Delta \tau},   \label{eq15} \end{equation}
that is to say, to send a message over a short time interval $\Delta \tau$, the message needs to be encoded at a high enough energy, the longer Bob stays outside collecting Hawking radiation, the less time Alice has to send her message, the more energy she needs to send this message. Alice encodes all the information in the message and send it to Bob, the required energy of this process is
\begin{equation} \Delta E_{H U P} \sim \frac{1}{G M} e^{G M^2},   \label{eq16} \end{equation}
which implies that the required energy for Alice is definitely larger than the black hole mass, i.e., $\Delta E \gg M$, therefore information must be encoded into the message with super-Planckian frequency, in other words, there is no violation of the no-cloning theorem.

 {\centering  \section{Black hole complementarity with higher order GUP} \label{secIII} }

The new higher order GUP is \cite{ret HA}
\begin{equation} [x, p]=\frac{i}{1-\alpha|p|}, \alpha>0   \label{eq17} \end{equation}
where $\alpha=\alpha_0 l_p=\frac{\alpha_0}{m_p}=\alpha_0 \sqrt{G}$($m_p$ is the Planck mass), this commutation relation contains a singularity at $|p|=\frac{1}{\alpha}$, and implies that the momentum of the particle cannot exceed $\frac{1}{\alpha}$, which agrees with Doubly Special Relativity \cite{ret WS}.

The uncertainty relation that arises from the higher order GUP is given by
\begin{equation} \Delta x \Delta p \geq \frac{1}{2}\left\langle\frac{1}{1-\alpha|p|}\right\rangle  =\frac{1}{2}\left\langle 1+\alpha|p|+\alpha^2 p^2+\alpha^3 p^2|p|+\alpha^4\left(p^2\right)^2+\cdots\right\rangle   \geq \frac{1}{2}\left(-\alpha \Delta p+\frac{1}{1-\alpha \Delta p}\right),   \label{eq18} \end{equation}
obviously, this formalism contains the different power function of uncertain momentum, which presents different algebraic structure from Pedram¡¯s one \cite{ret PP} and overcomes some conceptual problems raised in the original GUP\eqref{eq2} such as the divergence of the energy spectrum of the eigenfunctions of the position operator \cite{ret WSC}, that's exactly why we chose this form of GUP. In this paper we use this new formalism which has the maximal momentum $(\Delta p)_{max}<\frac{1}{\alpha}$ as well as the minimal length $(\Delta x)_{min}=\frac{3\alpha}{2}$. Next, by using the improved method, we calculate the modified temperature of the Schwarzschild black hole which is commensurate with the new order GUP. Then we transform the right-hand part of Eq.\eqref{eq18} into a perfect square trinomial of $\Delta p$ and take its square root to obtain
\begin{equation} \frac{1}{2 \alpha}\left(1-\sqrt{\frac{-3 \alpha+2 \Delta x}{\alpha+2 \Delta x}}\right) \leq \Delta p \leq \frac{1}{2 \alpha}\left(1+\sqrt{\frac{-3 \alpha+2 \Delta x}{\alpha+2 \Delta x}}\right),   \label{eq19} \end{equation}
analogy to Eq.\eqref{eq4}, we only choose the left-hand part of Eq.\eqref{eq19}, i.e.
\begin{equation} \frac{1}{2 \alpha}\left(1-\sqrt{\frac{-3 \alpha+2 \Delta x}{\alpha+2 \Delta x}}\right) \leq \Delta p, \label{eq20} \end{equation}
thus we have
\begin{equation} \left(\Delta p_{G 1}\right)_{\min }=\frac{1}{2 \alpha}\left(1-\sqrt{\frac{-3 \alpha+2 \Delta x}{\alpha+2 \Delta x}}\right), \label{eq21} \end{equation}
here if we consider $\Delta x \approx 2r_h = 4GM$ \cite{ret BMA}, we rewrite Eq.\eqref{eq21} as
\begin{equation}
\left(\Delta p_{G 1}\right)_{\min }=\frac{1}{2 \alpha}\left(1-\sqrt{\frac{-3 \alpha+8 G M}{\alpha+8 G M}}\right). \label{eq22} \end{equation}
The Heisenberg uncertainty principle $\Delta x \Delta p \geq \frac{1}{2}$ gives
\begin{equation}
\left(\Delta p\right)_{\min }=\frac{1}{2 \Delta x}, \label{eq23} \end{equation}
thus we can obtain an approximate expression for the revised differential of the area $d A_{G 1}$ as
\begin{equation}
d A_{G 1} \approx \frac{\left(\Delta p_{G 1}\right)_{\min }}{(\Delta p)_{\min }} d A \approx \frac{128 \pi G^3 M^2}{\alpha}\left(1-\sqrt{\frac{-3 \alpha+8 G M}{\alpha+8 G M}}\right) d M,
\label{eq24} \end{equation}
according to the Bekenstein-Hawking area law $S=\frac{A}{4{l_p}^2}$, Eq.\eqref{eq24} should be rewritten as
\begin{equation}
d S_{G 1} \approx \frac{32 \pi G^3 M^2}{\alpha l_p^2}\left(1-\sqrt{\frac{-3 \alpha+8 G M}{\alpha+8 G M}}\right) d M. \label{eq25} \end{equation}
where $d S_{G 1}$ is the differential of black hole entropy obtained from the improved method.

Now, by using the improved method, we express the corrected temperature of Schwarzschild black hole as
\begin{equation}
 T_{G 1}=\frac{d M}{d S_{G 1}}=\frac{(\alpha+8 G M) l_p^2}{128 \pi G^3 M^2}\left(1+\sqrt{\frac{-3 \alpha+8 G M}{\alpha+8 G M}}\right). \label{eq26} \end{equation}
Next, we focus on the required energy (GUP-corrected) for Alice. For oscillators in the box, the energy density is written in an integral form as
\begin{equation}
\rho=\frac{1}{V}  \int \overline{E} g(v)dv=2\int \frac{E}{e^{\frac{E}{T}}-1} d^3v, \label{eq27} \end{equation}
these photons will satisfy a GUP-corrected de Broglie relation which in the light of Eq. \eqref{eq18}, their average wavelength $lambda$ set equal to the minimum uncertainty on position, i.e., $(\Delta x) \min =\frac{1}{2}\left(-\alpha+\frac{1}{\Delta p(1-\alpha \Delta p)}\right)$, now reads
\begin{equation} \lambda=\frac{1}{2}\left(-\alpha+\frac{1}{p(1-\alpha p)}\right), \label{eq28} \end{equation}.
Obviously, such de Broglie waves $(E = p)$ the frequency can be written as
\begin{equation} v=\frac{1}{\lambda}, \label{eq29} \end{equation}
according to Eq.\eqref{eq18}, the range of the energy is $0 \leq E <\frac{1}{\alpha}$. In addition, from Eq.\eqref{eq29}, it is easily seen that
\begin{equation} d^3 v \approx\left[8 \pi E^2\left(1-5 E^2 \alpha^2-6 E^3 \alpha^3+7 E^4 \alpha^4\right)+O\left(\alpha^5\right)\right] d E.  \label{eq30} \end{equation}
Therefore, the energy density at a given temperature reads
\begin{equation}
\begin{split}
 \rho\left(T_{G 1}\right)=2 \int \frac{E}{e^{\frac{E}{T_{G 1}}}-1} d^3 v \approx 16 \pi \int \frac{\xi^3 T_{G 1}^4}{e^{\xi}-1}\left(1-5 \xi^2 T_{G 1}^2 \alpha^2-6 \xi^3 T_{G 1}^3 \alpha^3+7 \xi^4 T_{G 1}^4 \alpha^4\right) d \xi \\ \approx 16 \pi\left(\frac{\pi^4 T_{G 1}^4}{15}-5 \alpha^2 T_{G 1}^6\left(\frac{8 \pi^6}{63}\right)-6 T_{G 1}^7 \alpha^3\left(\frac{720 \pi^7}{2995}\right)+7 T_{G 1}^8 \alpha^4\left(\frac{8 \pi^8}{15}\right)\right).  \end{split}
 \label{eq31} \end{equation}

Furthermore, the GUP-corrected Stefan-Boltzmann law is
\begin{equation} \frac{d M}{d t} \approx 16 \pi G^2 M^2 \rho\left(T_{G 1}\right), \label{eq32} \end{equation}
thus Page time for the Schwarzschild black hole now becomes
\begin{equation} t_p(M) \approx \int \frac{1}{16 \pi G^2 M^2 \rho\left(T_{G 1}\right)} d M. \label{eq33} \end{equation}
Subsequently, according to the Ref.\cite{ret DN}, we rewrite the page time in terms of the black hole mass up to the fourth term for the large black hole as
\begin{equation} t_p \sim G^2 M^3+2 M \alpha^2-\frac{3 \alpha^4}{M G^2}+\frac{2}{G} \alpha^3 \log (M), \label{eq34} \end{equation}
thus Alice's proper time available to send Message should be
 \begin{equation} \Delta \tau \sim  G M \exp \left(-G M^2-\frac{2 \alpha^2}{G}+\frac{3 \alpha^4}{M^2 G^3}-\frac{2 \alpha^3}{G^2 M} \log (M)\right), \label{eq35} \end{equation}
besides, from Eq.\eqref{eq18}, it is easily seen that the proper energy-time uncertainty principle is
\begin{equation}
\Delta \tau \Delta E \geq \frac{1}{2}\left(-\alpha \Delta E+\frac{1}{1-\alpha \Delta E}\right). \label{eq36} \end{equation}
Now, Alice encodes all the information in the message and send it to Bob, the required energy of this process becomes
\begin{equation}
(\Delta E)_{G 1-\min }=\frac{1}{2 \alpha}\left(1-\sqrt{\frac{2 \Delta \tau-3 \alpha}{\alpha+2 \Delta \tau}}\right), \label{eq37} \end{equation}
we make a taylor expansion of Eq.\eqref{eq37} using the dimensionless GUP parameter $\alpha$ as
\begin{equation}
\begin{split}
 (\Delta E)_{G 1-\min } \sim \frac{e^{G M^2}}{2 G M}+\frac{e^{G M^2}\left(e^{2 G M^2}+8 G M^2\right) \alpha^2}{8 G^3 M^3}+\frac{e^{G M^2}\left(e^{3 G M^2}+16 G M^2 \log (M)\right) \alpha^3}{16 G^4 M^4}+\\ \frac{e^{G M^2}\left(3 e^{4 G M^2}+24 e^{2 G M^2} G M^2-48 G M^2+32 G^2 M^4\right) \alpha^4}{32 G^5 M^5}+O\left(\alpha^5\right).
 \end{split}   \label{eq38} \end{equation}

\begin{figure}[htbp]
\centerline{\includegraphics[width=7.0cm]{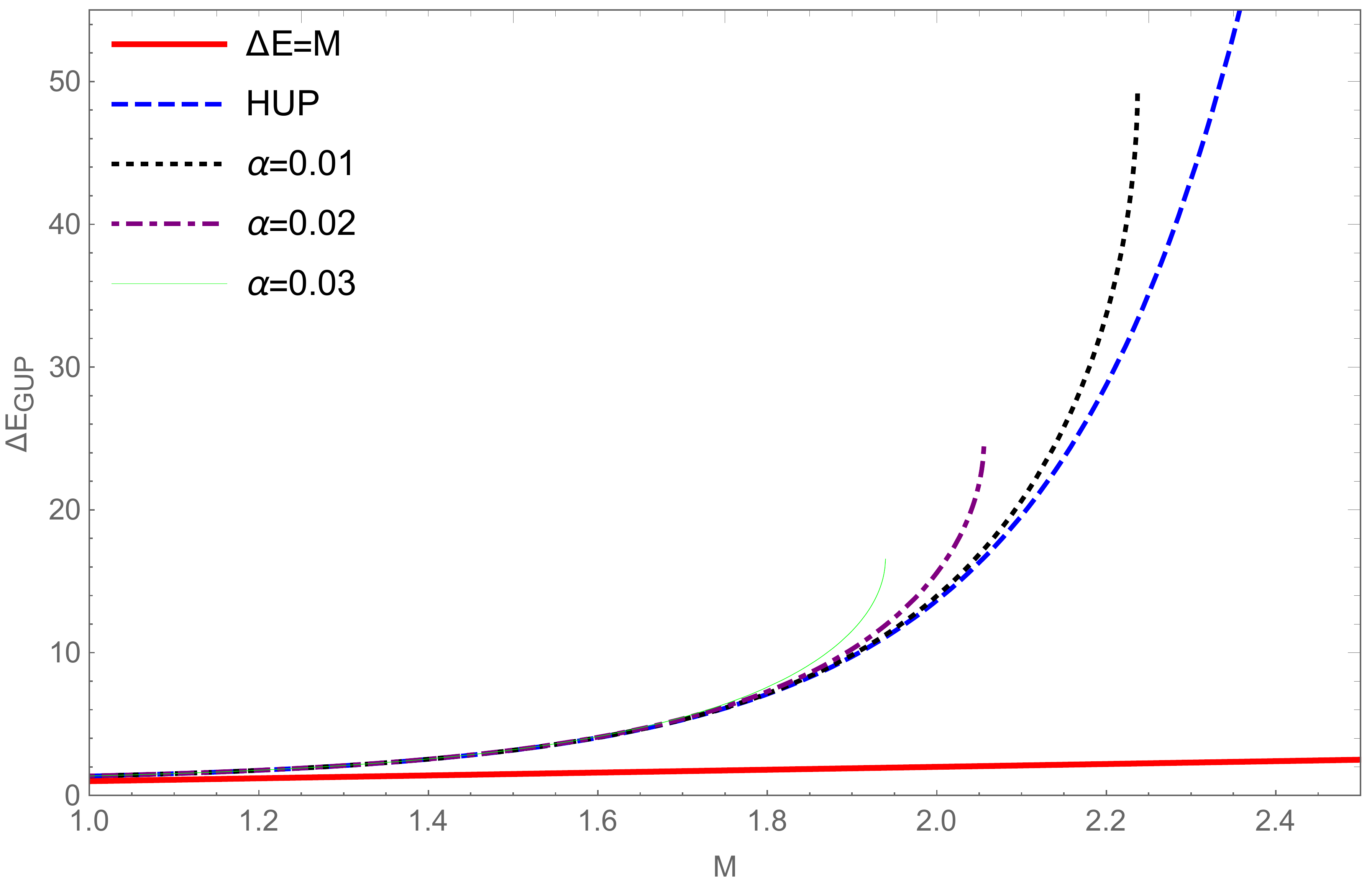}}
\caption{ In this figure, we have used Planck units, i.e., $G=c=\hbar=1$, so $l_p=1$, $m_p=1$.  The standard result £¨HUP£©and Eq.\eqref{eq37} for different GUP parameters are plotted, respectively. The red line describes the case in which the energy $\Delta E$ equals the black hole mass $M$.}
\label{fig1}
\end{figure}

The first term in the above equation is the uncorrected energy, by contrast with the Eq.\eqref{eq16}, the reason why we remain the coefficient $\frac{1}{2}$ is because we still had this constant coefficient  when we first thought about the GUP. The second term is the quadratic term of the corresponding GUP correction. The third term is the cubic term of the corresponding GUP correction. The fourth term is the fourth term of the corresponding GUP correction. In view of the fact that the GUP parameter is a small value, we will ignore the values after 4 times of GUP correction, which will not affect the scientific nature of the results. It shows that the GUP improve the no-cloning theorem, i.e. the energy required for a given black hole mass is greater than the required energy  for a black hole mass without GUP correction $(\Delta E)_{G-\min } \geq (\Delta E)_{HUP }>> M$, therefore, we conclude that black hole complementarity is still valid even in the presence of the new high order GUP.

{\centering  \section{recalculation} \label{secIV} }

As stated before, regarding the Ref. \cite{ret YG} and Ref. \cite{ret EC}, the same type of physics problem leads to two distinct-different conclusions. In this section, following the analysis in these two references, by using the improved method, we recalculate the corresponding GUP-corrected Hawking temperature. Noticed that the computational process of the GUP-corrected Hawking temperature and the Page time and the required energy in the context of GUP are similar with the Sec.\ref{secIII}, for the sake of convenience, we adopt the same approximate treatment as the original literatures and just express the corresponding results.

{\centering  \subsection{ Black hole complementarity with quadratic GUP}}

As shown in the Ref. \cite{ret YG}, the GUP with a quadratic term in momentum is
\begin{equation}
\Delta x \Delta p \geq  1+\alpha l_p^2 \Delta p^2,   \label{eq39}     \end{equation}
analogously, we transform the right-hand part of Eq.\eqref{eq39} into a perfect square trinomial of $\Delta p$ and take its square root to obtain:
\begin{equation}
\frac{\Delta x }{2\alpha l_p^2}\left[1+\sqrt{1-\frac{4\alpha l_p^2}{(\Delta x)^2}}\right] \geq \Delta p \geq \frac{\Delta x }{2\alpha l_p{ }^2}\left[1-\sqrt{1-4\frac{\alpha l_p{ }^2}{(\Delta x)^2}}\right],       \label{eq40}   \end{equation}
we choose the right-hand part of Eq.\eqref{eq40}, i.e. $\frac{\Delta x }{2\alpha l_p^2}\left[1-\sqrt{1-\frac{4\alpha l_p^2}{(\Delta x)^2}}\right] \leq \Delta p$, thus we have
\begin{equation} \left(\Delta p_{G2}\right)_{\min }=\frac{\Delta x }{2\alpha l_p^2}\left[1-\sqrt{1-\frac{4\alpha l_p^2}{(\Delta x)^2}}\right], \label{eq41} \end{equation}
also assuming that $\Delta x \approx 2r_h = 4GM$ \cite{ret BMA}, we transform Eq.\eqref{eq41} into
\begin{equation} \left(\Delta p_{G2}\right)_{\min }\approx \frac{2GM }{\alpha l_p^2}\left[1-\sqrt{1-\frac{\alpha l_p^2}{4G^2 M^2}}\right]. \label{eq42} \end{equation}

For the sake of convenience, we express the corrected temperature of Schwarzschild black hole by using the improved method as
 \begin{equation}
 T_{G 2}=\frac{d M}{d S_{G 2}}=\frac{l_p^2}{16 \pi G^2 M}\left(1+\sqrt{1-\frac{l_p^2 \alpha}{4 G^2 M^2}}\right), \label{eq43} \end{equation}
 and the page time becomes
\begin{equation}  t_p \sim G^2 M^3+G M \alpha,  \label{eq44} \end{equation}
and Alice¡¯s proper time available to send Message is
\begin{equation} \Delta \tau \sim r_h \exp \left(-\frac{t_p}{r_h}\right) \sim G M \exp \left(-G M^2-\alpha\right), \label{eq45} \end{equation}
thus Alice encodes all the information in the message and send it to Bob, the required energy of the process reads
\begin{equation}
(\Delta E)_{G 2-\min }=\frac{\Delta \tau}{2 l_p^2 \alpha}\left(1-\sqrt{1-\frac{4 \alpha l_p^2}{\Delta \tau^2}}\right) \sim \frac{M}{2 \alpha} e^{-\left(G M^2+\alpha\right)}\left(1-\sqrt{1-\frac{4 G \alpha}{\left.G^2 M^2 e^{-2\left(G M^2+\alpha\right.}\right)}}\right), \label{eq46} \end{equation}
analogously, we make a taylor expansion of Eq.\eqref{eq46} using the dimensionless GUP parameter $\alpha$ as
\begin{equation}
(\Delta E)_{G 2-\min } \sim \frac{e^{G M^2}}{G M}+\frac{e^{G M^2}\left(e^{2 G M^2}+G M^2\right) \alpha}{G^2 M^3}. \label{eq47} \end{equation}

\begin{figure}[htbp]
\centerline{\includegraphics[width=7.0cm]{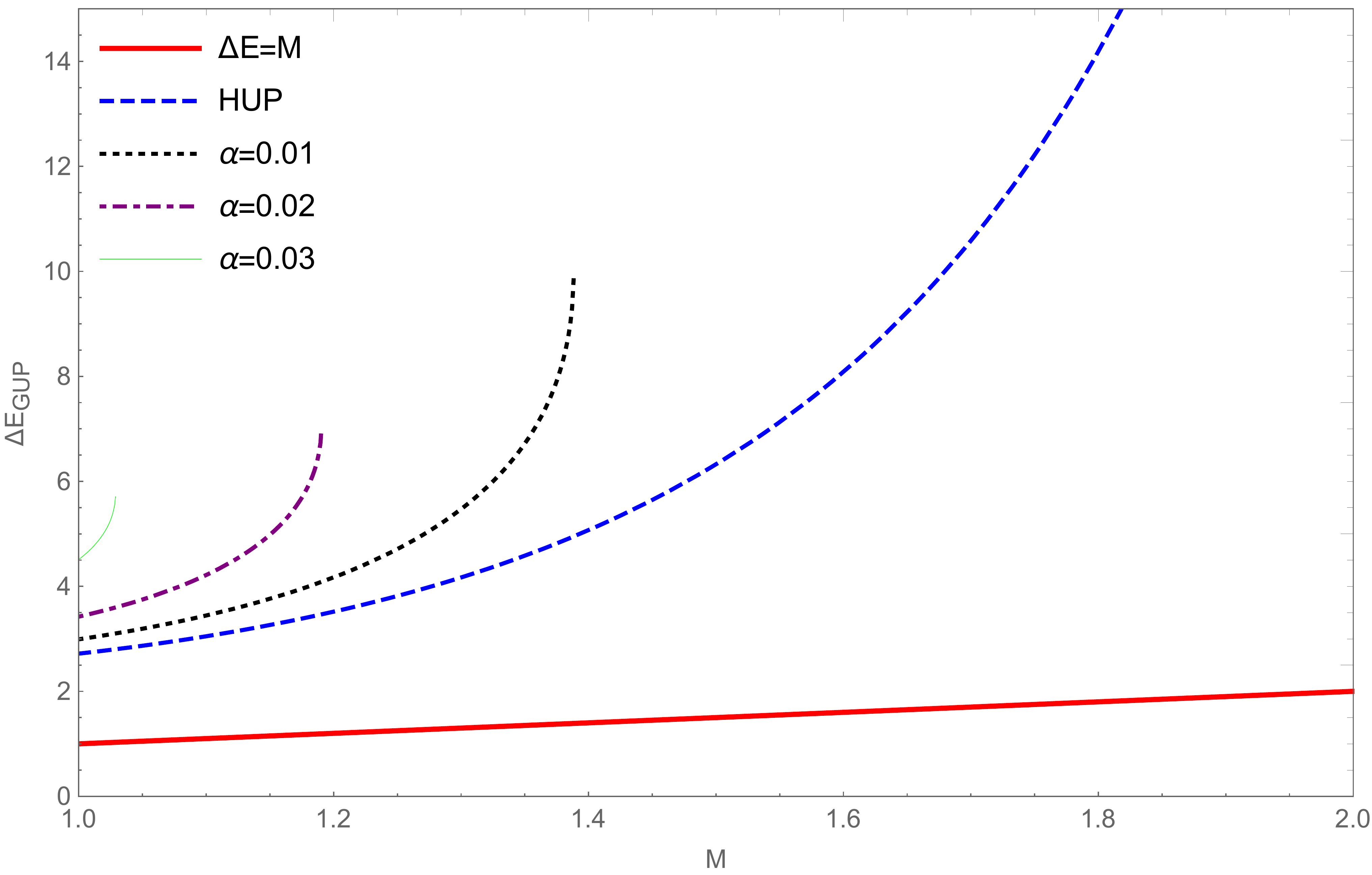}}
\caption{ In this figure, we have used Planck units, i.e., $G=c=\hbar=1$, so $l_p=1$, $m_p=1$.  The standard result £¨HUP£©and Eq.\eqref{eq46} for different GUP parameters are plotted, respectively. The red line describes the case in which the energy $\Delta E$ equals the black hole mass $M$.}
\label{fig2}
\end{figure}

The first term in Eq.\eqref{eq47} is the uncorrected energy, obviouly, this term is consistent with Eq.\eqref{eq16} when $\alpha\rightarrow0$.  The second term is the first term of the corresponding GUP correction. It shows that the GUP also improve the no-cloning theorem, i.e. the energy required for a given black hole mass is greater than the required energy  for a black hole mass without GUP correction $(\Delta E)_{G2-\min } \geq (\Delta E)_{HUP }>> M$, thus the black hole complementarity is still valid even in the presence of the GUP.

{\centering  \subsection{Black hole complementarity with Linear and quadratic GUP}}
In the Ref. \cite{ret EC}, the GUP with linear and quadratic GUP is
\begin{equation}
\Delta x \Delta p \geq\left[1-\alpha \Delta p+4 \alpha^2(\Delta p)^2\right],  \label{eq48} \end{equation}
here analogy to Eq.\eqref{eq40}, we have
\begin{equation}
\frac{(\alpha+\Delta x)}{8 \alpha^2}\left(1-\sqrt{1-\frac{16 \alpha^2}{(\Delta x+\alpha)^2}}\right) \leq \Delta p \leq \frac{(\alpha+\Delta x)}{8 \alpha^2}\left(1+\sqrt{1-\frac{16 \alpha^2}{(\Delta x+\alpha)^2}}\right),  \label{eq49} \end{equation}
also we choose the left-hand part of Eq.\eqref{eq49} as
\begin{equation}
\frac{(\alpha+\Delta x)}{8 \alpha^2}\left(1-\sqrt{1-\frac{16 \alpha^2}{(\Delta x+\alpha)^2}}\right) \leq \Delta p.   \label{eq50} \end{equation}
By using the improved method, it is obvious that the temperature of the black hole will no longer be the Hawking temperature but a modified one as
\begin{equation}
T_{G 3}=\frac{d M}{d S_{G 3}}=\frac{(\alpha+4 G M) l_p^2}{64 \pi G^3 M^2}\left(1+\sqrt{1-\frac{16 \alpha^2}{(4 G M+\alpha)^2}}\right), \label{eq51} \end{equation}
and the page time turns out to be
\begin{equation} t_p \sim G^2 M^3-G M^2 \alpha+M \alpha^2, \label{eq52} \end{equation}
thus Alices proper time available to send Message becomes
\begin{equation} \Delta \tau \sim r_h \exp \left(-\frac{t_p}{r_h}\right) \sim G M \exp \left(-G M^2+M \alpha-\frac{\alpha^2}{G}\right), \label{eq53} \end{equation}
therefore, in this case, the required energy is
\begin{equation}\begin{split}
\left.(\Delta E)_{G 3-\min }=\frac{(\alpha+\Delta \tau)}{8 \alpha^2}\left(1-\sqrt{1-\frac{16 \alpha^2}{(\Delta \tau+\alpha)^2}}\right)  \sim  \frac{\left(\alpha+G M e^{-G M^2+M \alpha-\frac{\alpha^2}{G}}\right)}{8 \alpha^2}\right( 1-\sqrt{\left.1-\frac{16 \alpha^2}{\left(G M e^{-G M^2+M \alpha-\frac{\alpha^2}{G}}+\alpha\right)^2}\right)},\end{split}
\label{eq54} \end{equation}
we still make a taylor expansion of Eq.\eqref{eq54} using the dimensionless GUP parameter $\alpha$ as
\begin{equation}
\begin{split}
(\Delta E)_{G 3-\min } \sim \frac{e^{G M^2}}{G M}-\frac{e^{G M^2}\left(e^{G M^2}+G M^2\right) \alpha}{G^2 M^2} \\+ \frac{e^{G M^2}\left(10 e^{2 G M^2} +2 G M^2+4 e^{G M^2} G M^2+G^2 M^4\right) \alpha^2}{2 G^3 M^3}+O\left(\alpha^3\right). \end{split}  \label{eq55}  \end{equation}

\begin{figure}[htbp]
\centerline{\includegraphics[width=7.0cm]{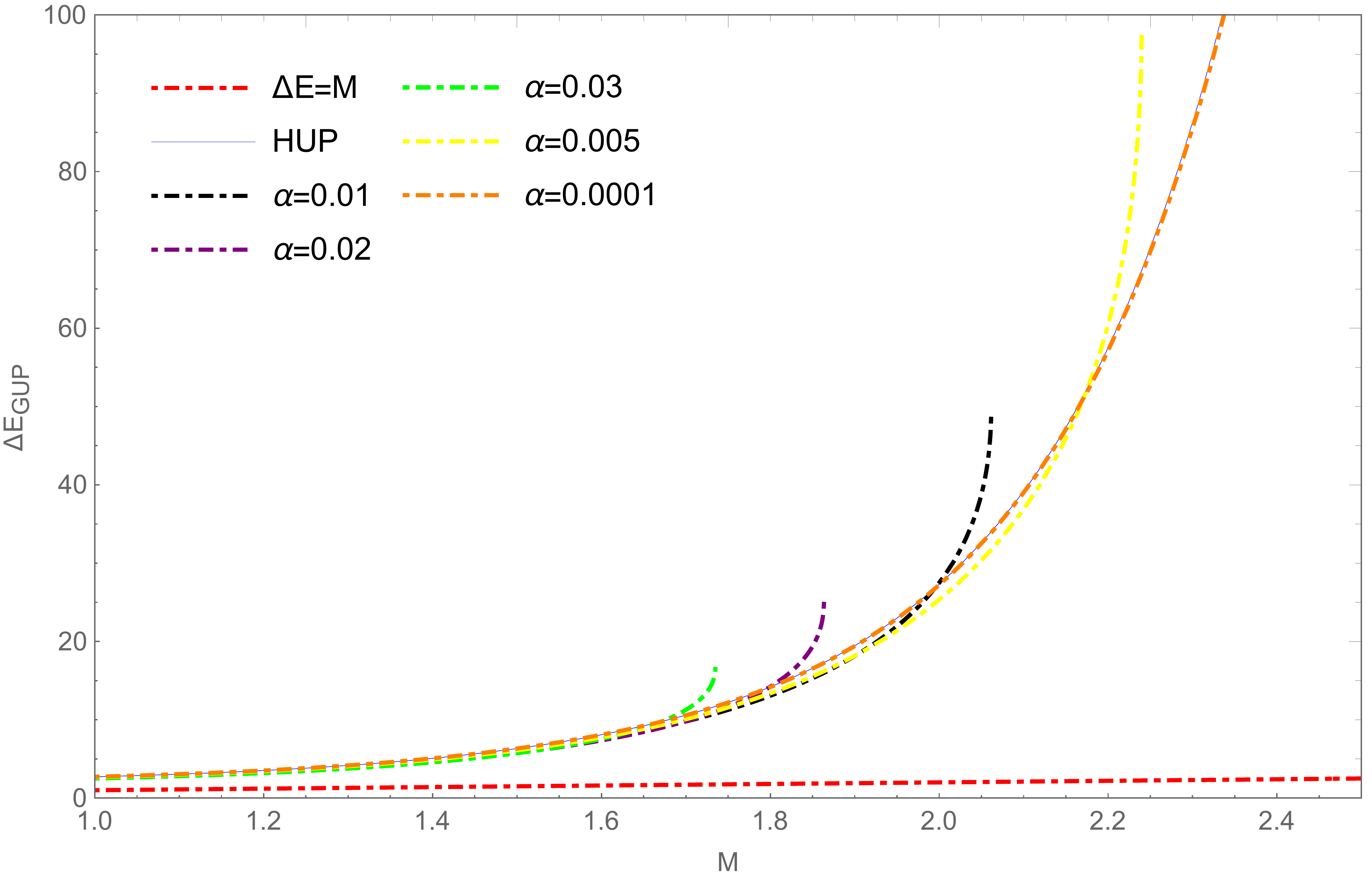}}
\caption{In this figure, we have used Planck units, i.e., $G=c=\hbar=1$, so $l_p=1$, $m_p=1$.  The standard result £¨HUP£©and Eq.\eqref{eq54} for different GUP parameters are plotted, respectively. The red line describes the case in which the energy $\Delta E$ equals the black hole mass $M$.)}
\label{fig3}
\end{figure}

The first term in Eq.\eqref{eq55} is the uncorrected energy, analogousy, this term is also consistent with Eq.\eqref{eq16} when $\alpha\rightarrow0$. The second term is the first term of the corresponding GUP correction. The third term is the second term of the corresponding GUP correction. Obviously, in this situation, the energy required for a given black hole mass is less than the required energy for a black hole mass with GUP correction $(\Delta E)_{G3-\min } \geq  M$, so according to the Stefan¨CBoltzmann law, black hole complementarity is also valid at an appropriate temperature.

{\centering  \section{conclusions} \label{secV} }

The core idea of this article is by using the improved method, the corrected Hawking temperature in the presence of different forms of GUP and the required energy for Alice to encode information in a message and send it to Bob are calculated, we conclude that for the three cases, the Black hole complementarity is always safe, this result present an answer for the conflicting results reported by Elias C. Vagenas et al and Yongwan Gim et al, and obviously, our result is reassuring, which also implies that the improved method is more reasonable in studying the thermodynamics of Schwarzschild black holes. In order to visualize our results, we present three figures according to Eq.\eqref{eq37}, Eq.\eqref{eq46} and Eq.\eqref{eq54}. From these figures, for the sake of comparison with Ref. \cite{ret YG} and Ref. \cite{ret EC}, we set the same parameters, i.e. $G=c=\hbar=l_p=m_p=1$. Noticed that we do not depict the curve for $\alpha=\sqrt{0.02}$ in Fig.\ref{fig3}, the reason for this behavior can be analyzed from Eq.\eqref{eq54}: since $\sqrt{1-\frac{16 \alpha^2}{\left(G M e^{-G M^2+M \alpha-\frac{\alpha^2}{G}}+\alpha\right)^2}} \geq 0$, when $G=1$ and $\alpha=\sqrt{0.02}$, the range of $M$ is $0.533575 \leq M \leq 0.974158$($M \geq m_p=1$). Naturally, when we take the parameters to be 1 in Fig.\ref{fig3}, the curve for $\alpha=\sqrt{0.02}$ is not available.  Thus we obtain the following results:

(1)Due to the effect of GUP, there are upper bounds at $(\Delta E)_{GUP }$, and which can be verified in our calculation, i.e. there is square root term in Eq.\eqref{eq37}, Eq.\eqref{eq46} and Eq.\eqref{eq54}.

(2) All the curves ($\alpha \neq 0$) are above the line ($\Delta E=M$), it impies that Black hole complementarity is safe in the presence of GUP.

(3) The required energy $(\Delta E)_{GUP }$ will increase with the increasing GUP parameter $\alpha$, which shows that GUP improve the no-cloning theorem.

\begin{acknowledgments}
This research was funded by the Guizhou Provincial Science and Technology Project(Guizhou Scientific Foundation-ZK[2022] General 491), National Natural Science Foundation of China (Grant No.12265007 and the Doctoral Foundation of Zunyi Normal University of China under Grants No. BS [2022]07, QJJ-[2022]-314).
\end{acknowledgments}


\begin{thebibliography}{77}

 \bibitem{ret SW} S.~W.~Hawking, Commun. Math. Phys. \textbf{43}, 199 (1975).

 \bibitem{ret SWH} S.~W.~Hawking, Phys. Rev. D \textbf{14}, 2460 (1976).

  \bibitem{ret DN} D.~N.~Page, Phys. Rev. Lett. \textbf{71}, 3743 (1993).


\bibitem{ret LS} L. Susskind and L. Thorlacius, Phys. Rev. D \textbf{49}, 966 (1994).
\bibitem{ret LSU} L. Susskind, L. Thorlacius and J. Uglum, Phys. Rev. D \textbf{48}, 3743 (1993).

 \bibitem{ret AF} A.~F.~Ali, S.~Das and E.~C.~Vagenas, Phys. Lett. B \textbf{678}, 497 (2009).
 \bibitem{ret SD} S.~Das and E.~C.~Vagenas, Phys. Rev. Lett. \textbf{101}, 221301 (2008).
 \bibitem{ret RJ} R.~J.~Adler, P.~Chen and D.~I.~Santiago, Gen. Rel. Grav. \textbf{33}, 2101 (2001).
\bibitem{ret NI} N.~Itzhaki, Phys. Rev. D \textbf{54}, 1557 (1996).
\bibitem{ret PC} P.~Chen, Y.~C.~Ong and D.~h.~Yeom, JHEP \textbf{12}, 021 (2014).




 \bibitem{ret YG} Y.~Gim, H.~Um and W.~Kim, JCAP \textbf{02}, 060 (2018).

 \bibitem{ret EC} E.~C.~Vagenas, A.~Farag Ali and H.~Alshal, Eur. Phys. J. C \textbf{79}, 276 (2019).

 \bibitem{ret DU} X.~D.~Du and C.~Y.~Long, JCAP \textbf{04}, 031 (2022).


 \bibitem{ret DA} D.~Amati, M.~Ciafaloni and G.~Veneziano, Phys. Lett. B \textbf{216}, 41 (1989).

 \bibitem{ret HA} H.~Hassanabadi, E.~Maghsoodi and W.~S.~Chung, Eur. Phys. J. C \textbf{79}, 358 (2019).

 \bibitem{ret WS} W.~S.~Chung and H.~Hassanabadi, Phys. Lett. B \textbf{785}, 127 (2018).

 \bibitem{ret PP} P.~Pedram, Phys. Lett. B \textbf{714}, 317 (2012).

 \bibitem{ret WSC} W.~S.~Chung and H.~Hassanabadi, Eur. Phys. J. C \textbf{79}, 213 (2019).


 \bibitem{ret BMA} B.~Majumder, Phys. Lett. B \textbf{703}, 402 (2011).




\end{thebibliography}
\end {document}